\documentclass[twocolumn,english,aps,prb,floats,showpacs]{revtex4}
\usepackage[T1]{fontenc}
\usepackage[latin9]{inputenc}
\usepackage{babel}
\usepackage{amsmath}
\usepackage{amssymb}
\usepackage{graphicx}
\usepackage{esint}
\usepackage[unicode=true,
 bookmarks=true,bookmarksnumbered=false,bookmarksopen=false,
 breaklinks=false,pdfborder={0 0 1},backref=false,colorlinks=false]
 {hyperref}
\hypersetup{pdftitle={Reversal of magnetization of a single-domain magnetic particle by the ac field of time-dependent frequency},
 pdfauthor={Liufei Cai, D. A. Garanin, and E. M. Chudnovsky},
 pdfkeywords={magnetic particle reversal microwave}}

\makeatletter

\newcommand{\lyxdot}{.}

\@ifundefined{textcolor}{}
{%
 \definecolor{BLACK}{gray}{0}
 \definecolor{WHITE}{gray}{1}
 \definecolor{RED}{rgb}{1,0,0}
 \definecolor{GREEN}{rgb}{0,1,0}
 \definecolor{BLUE}{rgb}{0,0,1}
 \definecolor{CYAN}{cmyk}{1,0,0,0}
 \definecolor{MAGENTA}{cmyk}{0,1,0,0}
 \definecolor{YELLOW}{cmyk}{0,0,1,0}
}

\usepackage{babel}

\makeatother

\begin{document}
\bibliographystyle{prsty}

\title{Reversal of magnetization of a single-domain magnetic particle \\
by the ac field of time-dependent frequency}

\author{Liufei Cai, D. A. Garanin, and E. M. Chudnovsky}

\affiliation{Physics Department, Lehman College, City University of New York \\
 250 Bedford Park Boulevard West, Bronx, New York 10468-1589, USA}

\date{\today}
\begin{abstract}
We report numerical and analytical studies of the reversal of the
magnetic moment of a single-domain magnetic particle by a circularly
polarized ac field of time-dependent frequency. For the time-linear
frequency sweep, the phase diagrams are computed that illustrate the
dependence of the reversal on the frequency sweep rate $v$, the amplitude
of the ac field $h$, the magnetic anisotropy field $d$, and the
damping parameter $\alpha$. It is shown that the most efficient magnetization
reversal requires a non-linear time dependence of the frequency, $\omega(t)$,
for which an exact analytical formula is derived with account of damping.
The necessary condition of the reversal is $h>\alpha d$. Implementation
of a small-scale magnetization reversal is proposed in which a nanomagnet
is electromagnetically coupled to two weak superconducting links controlled
by the voltage. Dynamics of such a system is analyzed with account
of the back effect of the magnet on the superconducting links. 
\end{abstract}


\maketitle

\section{Introduction}

In recent years a significant effort has been made to achieve magnetization
reversal in nanostructures, assisted by the low amplitude ac field
in the radio frequency range. The idea is rather simple. The dc magnetic
field required to reverse the magnetization of a single-domain magnetic
particle, the so-called anisotropy field, is typically in the range
0.01-0.1T. The field of this strength at the location of the particle
is not easy to develop fast. The ac magnetic field that one can typically
develop in the radio-frequency range would be two orders of magnitude
weaker. Applied in a resonant fashion, it could increase the amplitude
of the precession of the magnetic moment, sometimes leading to its
full reversal, in the same way as weak pushes of a pendulum at the
frequency of its mechanical oscillation can flip the pendulum over
the top. However, the study of both problem shows a lack of robust
reversal. In many cases the attempted reversal results in a chaotic
behavior that may be interesting on its own.

Magnetization reversal by ultrashot magnetic field pulses produced
by a high-energy electron beam has been studied by Back et al. \cite{Back-PRL1998}
in perpendicularly magnetized CoPt films. Schumacher et al \cite{Schumacher-PRL2003}
studied phase coherent precessional magnetization reversal in spin
valves by a pulse of the transverse field of a few hundred picosecond
duration produced by the electric current. 

Later, significant number of experiments focused on microwave-assisted
reversal in smaller structures and individual single-domain magnets
with a strong static field applied to reduce the barrier. Thirion
et al \cite{Thirion-2003} attempted magnetization reversal in static
fields below the anisotropy field, assisted by a linearly polarized
microwave field, in $20$-nm-diameter Co particles placed on the bridge
of a micro-SQUID. They were able to reproduce the Stoner-Wohlfarth
astroid \cite{Lectures} and study the dependence of the reversal
on the frequency and duration of the ac pulse. Enhancement of the
magnetization reversal by microwave magnetic fields in nanometer Co
strips has been demonstrated by Grollier et al. \cite{Grollier-JAP2006}
Nembach et al \cite{Nembach-APL2007} and Nozaki et al \cite{Nozaki-APL2007}
used magnetic force microscopy to measure microwave assisted magnetization
reversal in individual submicron Co and permalloy particles. Microwave-assisted
magnetization switching in permalloy tunnel junctions has been demonstrated
by Moriyama et al \cite{Moriyama-APL2007}. Podbielski et al studied
magnetization reversal in microscopic permalloy rings at GHz frequency.
They observed non-linear spin dynamics and obtained experimental phase
diagram of the reversal as function of microwave frequency and power.
\cite{Podbielski-PRL2007} Using time-resolved magneto-optic Kerr
microscopy, Woltersdorf and Back \cite{Woltersdorf-PRL2007} detected
enhancement of magnetization switching in single-domain permalloy
elements subjected to the resonant microwave field. Microwave-assisted
magnetization reversal in single-domain permalloy nanoelements has
been studied by Nembach et al. \cite{Nembach-APL2009} Wang et al
\cite{Wang-PRB2010} have investigated experimentally the competition
between damping and pumping for microwave-assisted magnetization reversal
in FeCo thin films.

Theoretical research in this area mostly focused on the magnetization
reversal assisted by the ac field of constant frequency. \cite{Bertotti-JAP2009}
Non-linear magnetization dynamics induced by such a field that results
in a chaotic behavior has been studied by Bertotti et al. \cite{Bertotti-PRL2001,Bertotti-JAP2002}
Denisov et al \cite{Denisov-PRL2006} addressed magnetization of nanoparticles
in a rotating magnetic field. Synchronization and chaos induced in
the damped dynamics of a single-domain particle by the ac field of
constant frequency has been investigated by Sun and Wang. \cite{Sun-PRB2006}
Nonlinear-dynamical-system approach to the microwave-assisted magnetization
dynamics was reviewed by Bertotti et al. \cite{Bertotti-JAP2009}
Micromagnetic modelling of microwave-assisted magnetic recording was
performed by Wang et al. \cite{Wang-JAP2009} Constant frequency microwave
switching magnetic grains coupled by exchange interaction has been
investigated by Igarashi et al. \cite{Igarashi-JAP2009} Okamoto et
al addressed stability of the magnetization switching by linearly
and circularly polarized waves. Magnetization reversal in a resonant
cavity has been studied by Yukalov and Yukalova. \cite{Yukalov-JAP2012}

Fewer number of theoretical papers have considered dynamics of the
magnetization of a nanoparticle generated by the ac magnetic field
of variable frequency. Mayergoltz et al \cite{Mayergoyz-JAP2004}
developed the inverse problem approach to the precessional switching
of the magnetization by a linearly polarized pulse of the magnetic
field. Rivkin and Ketterson \cite{Rivkin-APL2006} obtained the optimal
time dependence of the microwave frequency in the absence of damping,
as well as the condition of the reversal in the presence of damping.
Magnetization reversal by a linearly polarized ac field of frequency
that depends linearly on time has been studied by two of the authors.
\cite{LC-PRB2010} Barros et al \cite{Barros-PRB2011} developed an
optimization method in which the energy consumption needed for reversal
is minimized with respect to the time dependence of the amplitude
and frequency of microwaves.

A few general points need to be made before addressing the problem
of the reversal of the magnetization by the microwaves. Firstly, a
robust magnetization reversal can be effectively achieved only with
a circularly polarized ac field. Indeed, photons of circular polarization
have a definite orientation of their spin projection, while photons
with linear polarization are in a superposition of spin states. Consequently,
photons with the right circular polarization, when absorbed by the
magnet, drive the magnetization in one direction towards the reversal,
while linearly polarized photons can be both absorbed and emitted
and, therefore, do not necessarily reverse the magnetization. Secondly,
the photons are effectively absorbed only when they are in resonance
with the spin levels. The latter are not equidistant on the magnetic
quantum number, that is, on the projection of the magnetic moment
on the direction of the effective field. Thus, as the magnetic moment
reverses, the photon frequency that can be resonantly absorbed by
the magnet changes with time, so that the frequency of the microwave
field has to be adjusted. Damping of the precession adds another dimension
to this problem as the power of the ac field that is pumped into the
magnet should exceed the rate of energy dissipation. Analysis shows
that circularly polarized small-amplitude ac field of a time-dependent
frequency that follows the condition of the resonance is sufficient
for achieving magnetization reversal. The case of a zero static field
is of the highest practical importance.

The typical wavelength of microwaves that are in resonance with the
precession of the magnetic moment is in the centimeter range. Thus,
one of the challenges for potential applications of the microwave-assisted
magnetization reversal for, e.g., computer technology, consists of
the generation of a circularly polarized ac field of sufficient amplitude
at the position of a nanoscale single-domain particle. In Ref.\ \onlinecite{LC-PRB2010}
a suggestion has been made to use the ac field generated by a superconducting
weak link. If one is not turned off by the necessity to go to lower
temperatures (which is probably, inevitable for magnetic memory of
ultra-high density), the advantage of this method would be the possibility
to control the time dependence of the frequency by voltage across
the link. Interaction between a nanomagnet and a Josephson junction
has been subject of intensive research. Micro-SQUID setup has been
used by Jamet et al to observe switching of the magnetization in a
3nm Co cluster \cite{Jamet-PRL2001,Wernsdorfer-2001}, see also review
\onlinecite{Wernsdorfer-2009}. Ferromagnetic resonance in permalloy
films grown on Nb substrate has been studied by Bell et al. \cite{Bell-2008}
Petcovi\'{c} et al \cite{Petkovic-2009} investigated experimentally
the direct dynamical coupling of spin modes and a supercurrent in
a ferromagnetic junction, following theoretical study of this system
by Houzet. \cite{Houzet-2008} Current-phase relation in a Josephson
junction coupled with a magnetic dot has been investigated theoretically
by Samokhvalov. \cite{Samokhvalov-2009} Most of the research in this
area focused on the proximity effect \cite{Buzdin-review,Buzdin-2008,Buzdin-2009}
rather than on electromagnetic interaction.

In this paper we study magnetization dynamics of a single-domain uniaxial
magnetic particle in zero static field, induced by a circularly polarized
ac field of constant amplitude but variable frequency. The model is
formulated in Section \ref{model}. General properties of the magnetization
reversal are studied in Section \ref{general}. Numerical results
for the time-linear frequency sweep are presented in Section \ref{numerical-linear}
where the phase diagrams are computed for the dependence of the magnetization
switching on parameters. They are the frequency sweep rate, the amplitude
of the ac field, the magnetic anisotropy field, and the damping parameter.
Analytical results for the time-linear sweep, that are generally in
good agreement with numerical results, are given in Section \ref{analytical-linear}.
In Section \ref{optimal} we obtain the exact analytical solution
for non-linear time dependence of the frequency that provides the
fastest magnetization reversal. The model in which circularly polarized
ac field is generated by two superconducting weak links is studied
in Section \ref{JJ} with account of the back effect of the dynamics
of the magnetic moment on the links. Our conclusions and suggestions
for experiment are presented in Section \ref{conclusion}. 

\section{The model}

\label{model}

The energy of a single-domain magnetic particle with an uniaxial anisotropy
in a circularly-polarized ac field has the form 
\begin{equation}
\mathcal{H}=-KVM_{z}^{2}-VM_{x}h\cos\Phi(t)-VM_{y}h\sin\Phi(t).\label{eq:Ham-0}
\end{equation}
Here $K$ is the magnetic anisotropy constant, $V$ is particle's
volume, ${\bf M}$ is the magnetization, $h$ is the amplitude of
the ac field, and $\Phi(t)$ is the phase related to the time-dependent
frequency as 
\begin{equation}
\dot{\Phi}(t)\equiv\omega(t).\label{eq:Phase-Def}
\end{equation}
One of the cases we consider is that of the frequency linearly changing
with time, 
\begin{equation}
\omega(t)=-vt,\label{eq:omega_Def}
\end{equation}
where $\Phi(t)=-vt^{2}/2$. The other case that will be studied here
is a non-linear time dependence of the frequency that provides the
fastest magnetization reversal.

It is convenient to recast the problem in terms of a classical spin
$\mathbf{s}=\mathbf{M}/M_{s}$, $\left|\mathbf{s}\right|=1$, where
$M_{s}$ is the saturation magnetization. The Landau-Lifshitz equation
of motion for this spin has the form 
\begin{equation}
\mathbf{\dot{s}}=\gamma\left[\mathbf{s}\times\mathbf{H}_{\mathrm{eff}}\right]-\alpha\gamma\left[\mathbf{s}\times\left[\mathbf{s}\times\mathbf{H}_{\mathrm{eff}}\right]\right],\label{eq:LLE}
\end{equation}
where $\gamma$ is the gyromagnetic ratio, $\alpha$ is dimensionless
damping coefficient and 
\begin{equation}
\mathbf{H}_{\mathrm{eff}}=-\frac{1}{V}\frac{\partial\mathcal{H}}{\partial\mathbf{M}}=2ds_{z}\mathbf{e}_{z}+h\mathbf{e}_{x}\cos\Phi(t)+h\mathbf{e}_{y}\sin\Phi(t)\label{eq:h_eff_Def}
\end{equation}
is the effective field. Here $d\equiv H_{a}=KM_{s}$ is the anisotropy
field. In the initial state the spin points in the negative-$z$ direction,
$\mathbf{s}(-\infty)=-\mathbf{e}_{z}.$

Further it is convenient to switch to the coordinate frame rotating
around the $z$ axis together with the magnetic field, so that in
this frame the magnetic field is static. As the result, in the new
frame the spin acquires a rotation opposite to that of the ac field
in the initial (laboratory) frame. Thus in the rotating frame the
Landau-Lifshitz has the form 
\begin{equation}
\mathbf{\dot{s}}=\left[\mathbf{s}\times\left(\gamma\mathbf{H}_{\mathrm{eff}}+\mathbf{\Omega}(t)\right)\right]-\alpha\gamma\left[\mathbf{s}\times\left[\mathbf{s}\times\mathbf{H}_{\mathrm{eff}}\right]\right],\label{eq:LLE-Rotating_frame}
\end{equation}
where 
\begin{equation}
\mathbf{H}_{\mathrm{eff}}=2ds_{z}\mathbf{e}_{z}+h\mathbf{e}_{x},\qquad\mathbf{\Omega}(t)=\omega(t)\mathbf{e}_{z}.\label{eq:h_eff_omega_rotating_frame}
\end{equation}

\section{General properties of the magnetization reversal}

\label{general}

With the sign choice in Eq. (\ref{eq:omega_Def}), the ac field at
negative times is precessing in the same direction as the magnetic
moment, thus it excites magnetic resonance and may cause magnetization
reversal. In the ideal case, as we will see below, the resonance condition
holds during the whole reversal. After the magnetic moment overcomes
the barrier, $s_{z}>0$, it changes its precession direction, and
so does the ac field.

In the rotating frame, the field $\mathbf{\omega}(t)/\gamma$ sweeping
at a linear rate makes the problem resembling that of the Landau-Zener
(LZ) effect that can be formulated in terms of the evolution of a
classical spin described by a Larmor equation. There are three modifications,
however: (i) uniaxial anisotropy and (ii) damping added to the model
and (iii) sweeping $\mathbf{\omega}(t)$ in the negative direction.
Because of the latter, the initial state of the spin in the rotating
frame is the high-energy state with $s_{z}$ opposite to $\omega$,
see Eq. (\ref{eq:omega_Def}). To the contrast, in the regular LZ
effect the initial spin state is the low-energy state. In the absence
of anisotropy and damping, the initial orientation of the spin and
the sweep direction do not matter. However, in the general case the
situation does depend on these factors.

In particular, in the absence of damping one can multiply $\gamma\mathbf{h}_{\mathrm{eff}}+\mathbf{\Omega}(t)$
by $-1$ that only makes the spin precess in the opposite direction
but does not affect its reversal. The resulting model is a model with
a positive sweep, such as the regular LZ problem, while the anisotropy
becomes easy-plane, $d<0$. It was shown \cite{gar03prb} that in
this case for a sweep slow enough the system adiabatically follows
the time-dependent lowest-energy state and the spin switching is very
efficient. In our original model (with $\alpha=0$) the magnetization
reversal is similar. Only instead of adiabatically following the lowest-energy
state, the spin adiabatically follows the highest-energy state, in
which it was at the beginning.

This adiabatic solution corresponds to the maximum of the energy in
the rotating frame 
\begin{equation}
\mathcal{H}/(VM_{s})=-ds_{z}^{2}-s_{x}h-\left(\omega/\gamma\right)s_{z}\label{eq:Ham-Rot-frame}
\end{equation}
The maximal energy corresponds to $s_{y}=0$. Using $s_{x}=-\sqrt{1-s_{z}^{2}}$
(opposite to the transverse field) and requiring $d\mathcal{H}/ds_{z}=0$,
one obtains the equation 
\begin{equation}
2ds_{z}+hs_{z}/\sqrt{1-s_{z}^{2}}+\omega/\gamma=0\label{eq:Ham_max_Eq}
\end{equation}
for the energy maximum. Since in practical conditions $h\ll d$, an
approximate solution of this 4th-order algebraic equation for the
adiabatic spin value reads 
\begin{equation}
s_{z}=\begin{cases}
-1, & \omega>2\gamma d\\
-\omega/(2\gamma d), & \left|\omega\right|\leq2\gamma d\\
1, & \omega<-2\gamma d.
\end{cases}\label{eq:sz-Adi}
\end{equation}
Note that this solution is independent of $h$. Nonzero values of
$h$ cause rounding at the borders of the central region $\left|\omega\right|\leq2\gamma d$
where the reversal occurs. In the laboratory frame, the spin is precessing
during adiabatic reversal being phase-locked to the ac field.

For $\alpha=0$ the magnetization reversal can be achieved for whatever
small ac field $h$. In the case of a nonzero damping, there is a
dissipative torque acting towards the energy minima, and the ac field
$h$ has to exceed a threshold value to overcome this torque. Below
we will see that the magnetization reversal requires 
\begin{equation}
h>\alpha d\label{eq:h-alpha-condition}
\end{equation}
that is much easier to fulfill than $h>2d$ in the case of a static
field. As the torque due to the transverse field is maximal when the
magnetic moment is perpendicular to it, in the presence of damping
the magnetic moment goes out of the $x$-$z$ plane during the reversal.

\section{Numerical - magnetization reversal by the time-linear frequency sweep}

\label{numerical-linear}

\subsection{Time dependences of reversing magnetization}

\begin{figure}
\centering\includegraphics[width=8cm]{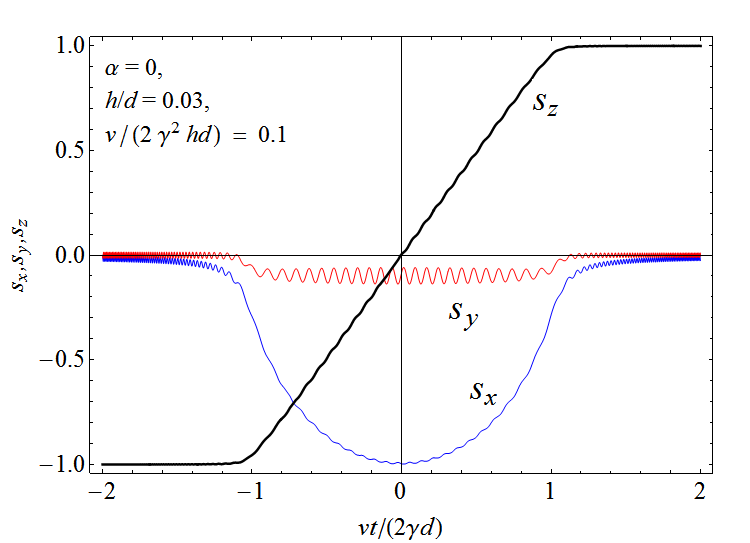}

\caption{Almost adiabatic magnetization reversal at zero damping.}

\label{Fig-spin_reversal-adiabatic} 
\end{figure}

\begin{figure}
\centering\includegraphics[width=8cm]{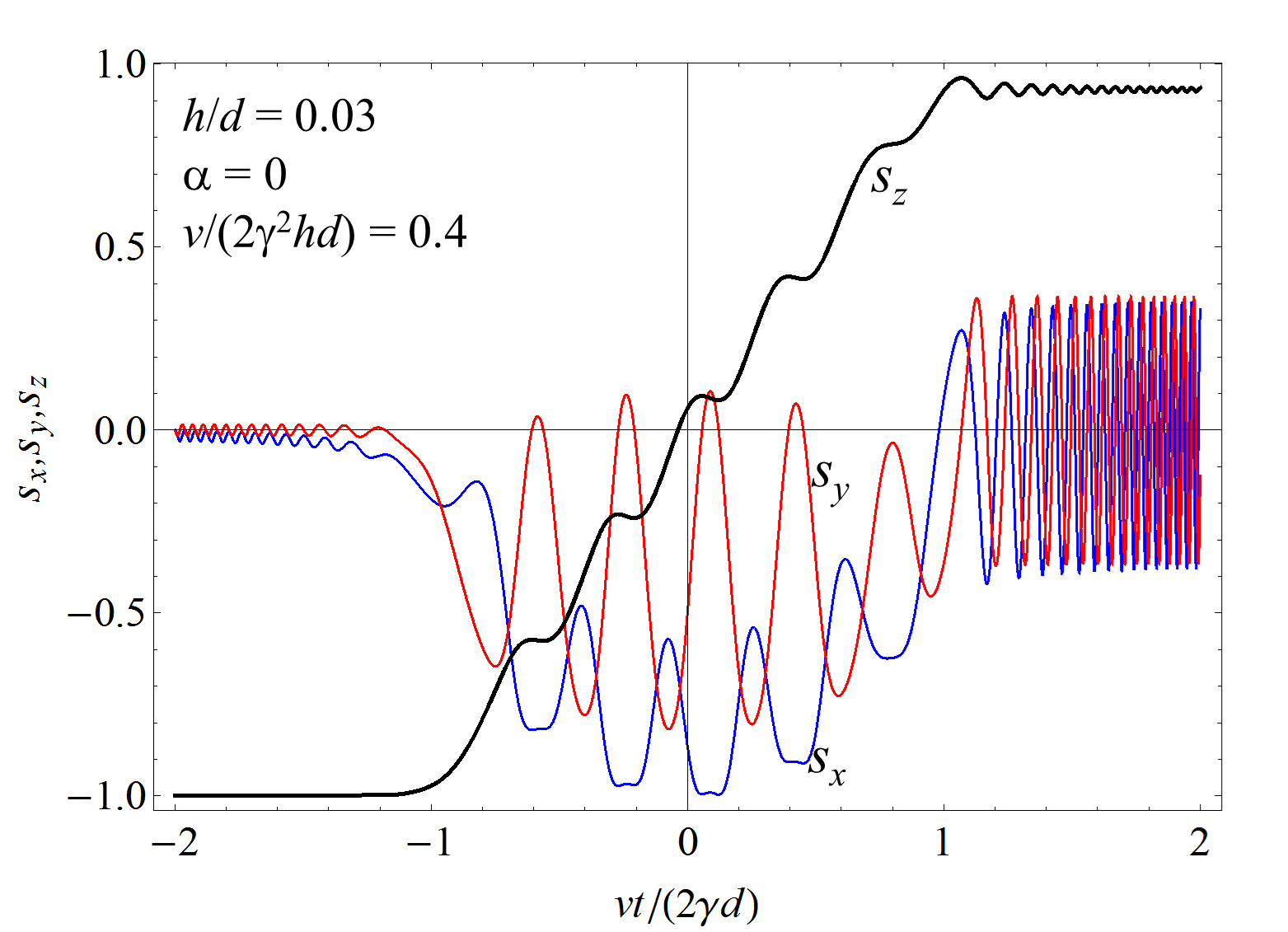}

\caption{Non-adiabatic magnetization reversal at zero damping.}

\label{Fig-spin_reversal-non-adiabatic} 
\end{figure}

The results of numerical solution of Eq. (\ref{eq:LLE-Rotating_frame})
in the undamped case $\alpha=0$ for a small frequency sweep rate
are shown in Fig. \ref{Fig-spin_reversal-adiabatic}. Magnetization
reversal in this case is almost adiabatic and $s_{z}(t)$ is well
described by Eq. (\ref{eq:sz-Adi}) with rounding at the borders of
the reversal interval due to a small value of $h/d$. The reversal
is practically confined to the $z$-$x$ plane and $s_{y}$ is small.
Numerical results for a faster sweep rate are shown in Fig. \ref{Fig-spin_reversal-non-adiabatic}.
Here there is still magnetization reversal but it is not adiabatic
and the final value of $s_{z}$ is smaller than one. Because of this,
the magnetic moment is precessing around the $z$ axis, as manifested
by $s_{x}$ and $s_{y}$. During reversal the magnetization is substantially
deviating from the $z$-$x$ plane. For larger sweep rates the reversal
quickly becomes impossible.

\begin{figure}
\centering\includegraphics[width=8cm]{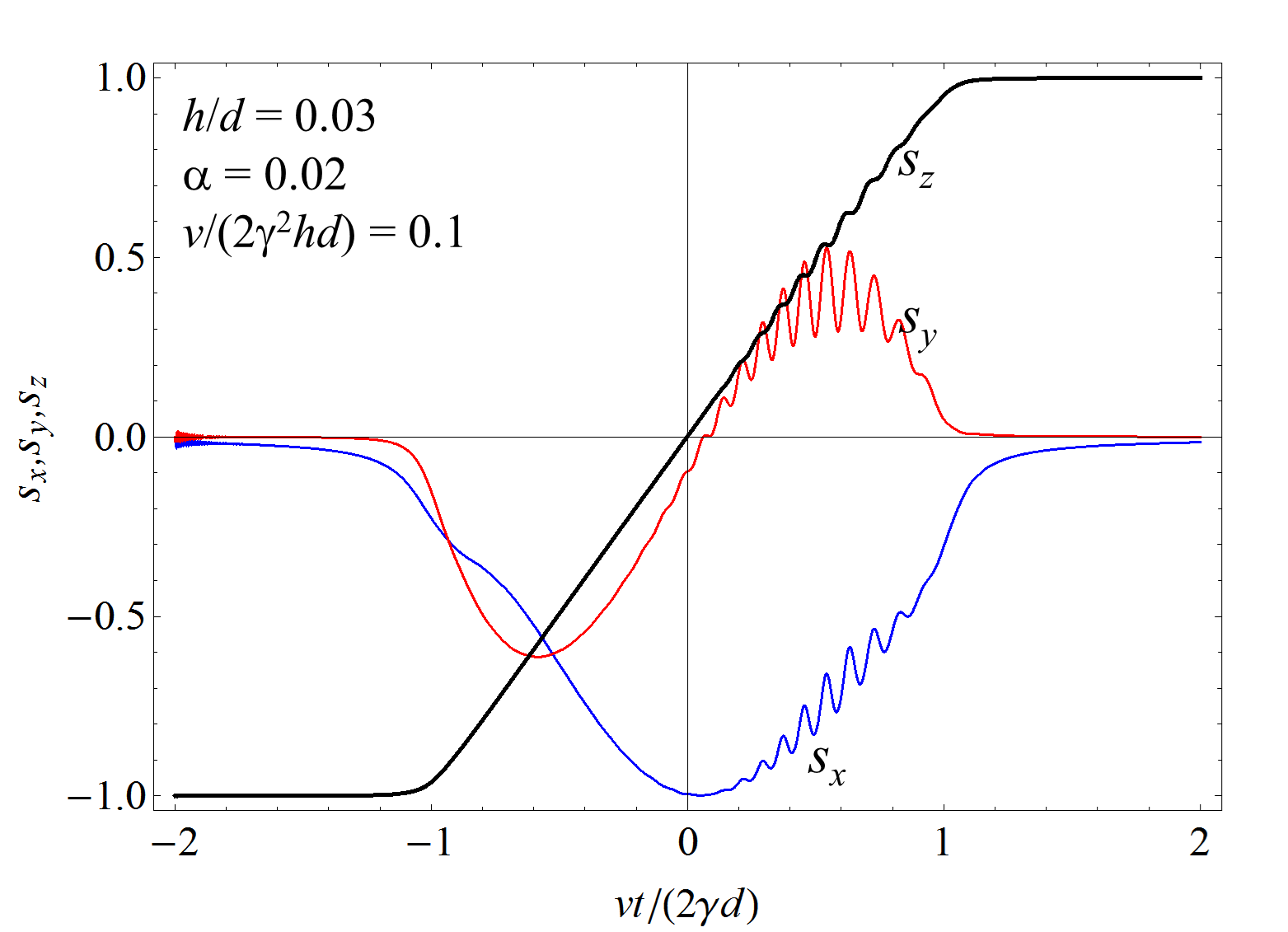}

\caption{Almost adiabatic magnetization reversal for $\alpha=0.02$.}

\label{Fig-spin_reversal-damped-adiabatic} 
\end{figure}

\begin{figure}
\centering\includegraphics[width=8cm]{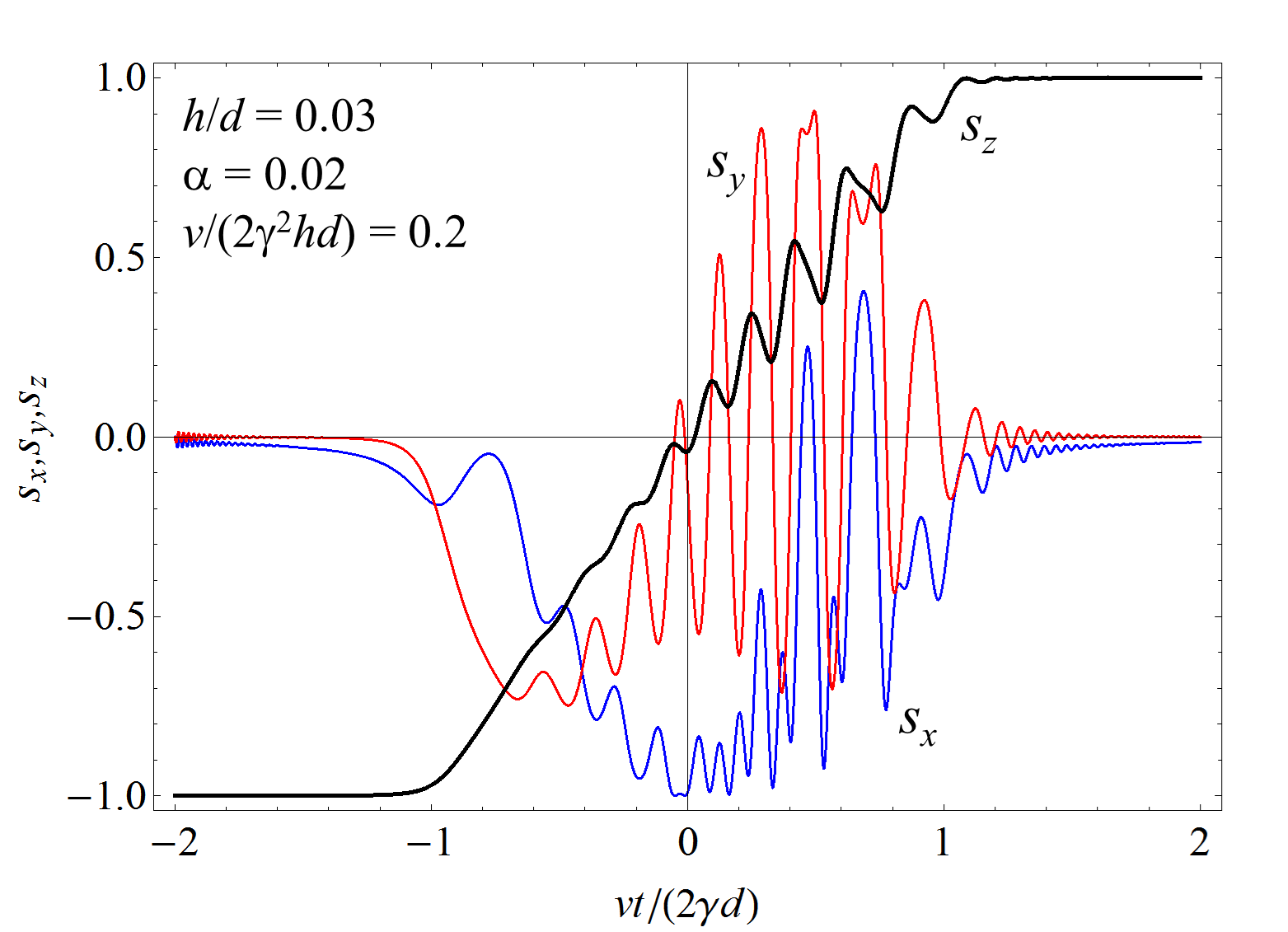}

\caption{Non-adiabatic magnetization reversal for $\alpha=0.02$.}

\label{Fig-spin_reversal-damped-non-adiabatic} 
\end{figure}

\begin{figure}
\centering\includegraphics[width=8cm]{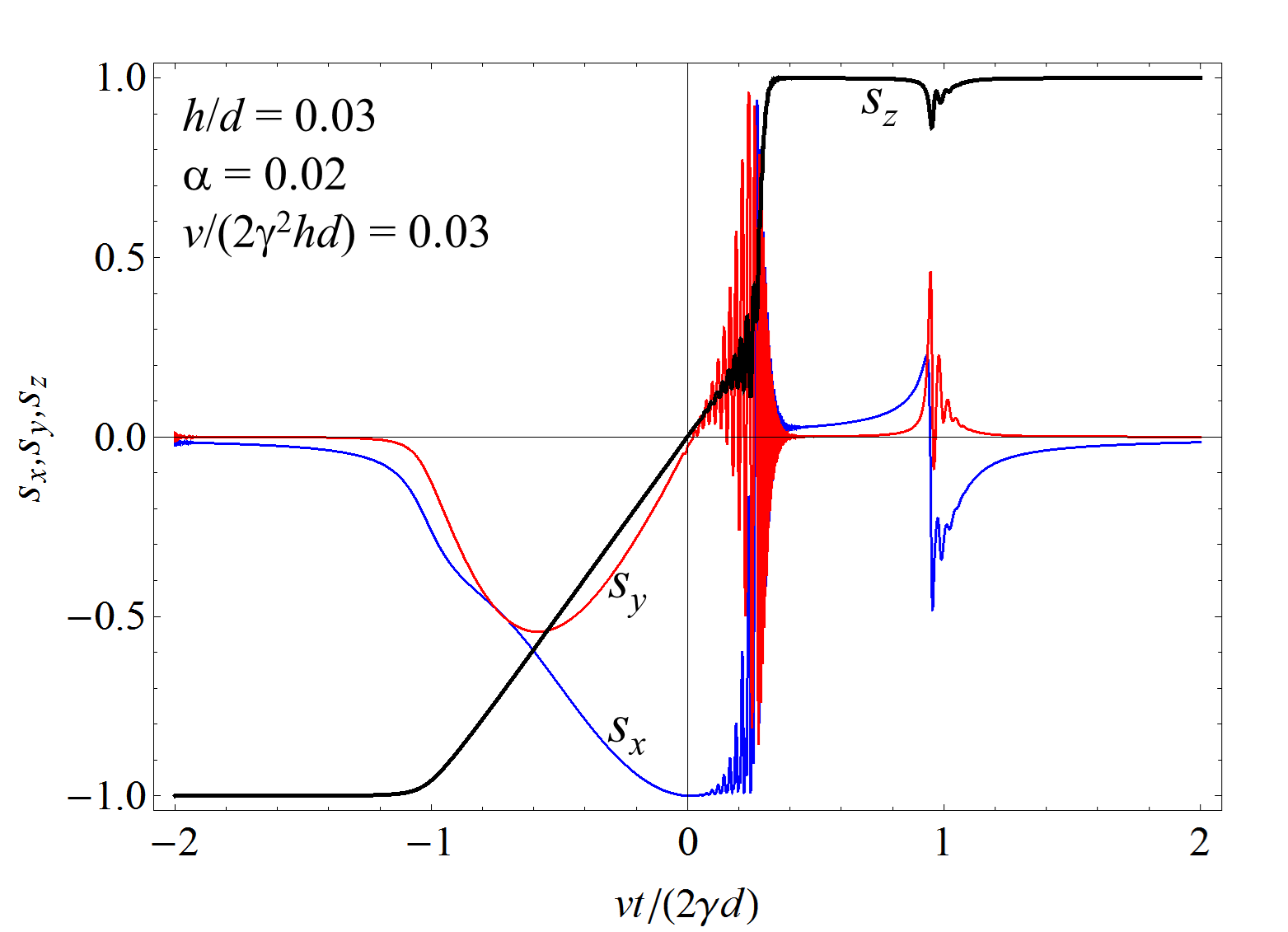}

\caption{Instability in slow magnetization reversal for $\alpha=0.02$.}

\label{Fig-spin_reversal-damped-slow} 
\end{figure}

Fig. \ref{Fig-spin_reversal-damped-adiabatic} shows that in the damped
case the magnetic moment substantially deviates from the $z$-$x$
plane. Still, overall the reversal in this case is close to adiabatic.
Increasing the sweep rate leads to a non-adiabatic regime shown in
Fig. \ref{Fig-spin_reversal-damped-non-adiabatic}. Here transverse
spin components are oscillating and the dependence of $s_{z}$ is
jagged. This shows that, in the laboratory frame, the phase locking
between the magnetic moment and the ac field is about to break. In
spite of all this, there is a complete reversal because the damping
finally brings the magnetic moment to the bottom of the potential
well (c.f. Fig. \ref{Fig-spin_reversal-non-adiabatic}). For a faster
sweep the reversal disappears and the magnetic moment lands in the
initial well, $s_{z}=-1$. In the case of a slow sweep shown in Fig.
\ref{Fig-spin_reversal-damped-slow} an instability can develop that
leads to the breakdown of the phase locking and to faster relaxation
of the magnetic moment towards one of the two potential wells. The
final value of $s_{z}$ (1 or -1) behaves irregularly vs sweep rate.
This regime is not interesting for applications aimed at achieving
as fast as possible reversal.

\subsection{Phase diagram of the magnetization reversal by the time-linear frequency
sweep}

\begin{figure}
\centering\includegraphics[width=8cm]{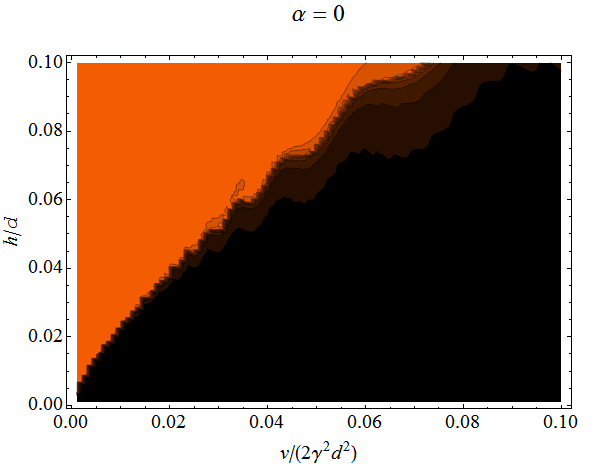}
\caption{Phase diagram of the magnetization reversal in the undamped case.}

\label{Fig-PD-indep-linear-h_alpha=00003D00003D0_100x100} 
\end{figure}

Dependence of the final value of $s_{z}$ on the amplitude of the
ac field $h$ and frequency sweep rate $v$ defines the ``phase diagram''
of the magnetization reversal. In the undamped case the numerically
calculated phase diagram is shown in Fig. \ref{Fig-PD-indep-linear-h_alpha=00003D00003D0_100x100}.
The final $s_{z}$ is color-coded: black corresponds to $s_{z}=-1$
(non-reversal) and red corresponds to $s_{z}=1$ (reversal). One can
see that the reversal requires $h$ sufficiently large and $v$ sufficiently
small. The curvature of the phase boundary at small $h$ and $v$
suggests a fractional power. Careful examenation of this region of
the phase diagram shows that the reversal condition has the form 
\begin{equation}
\frac{v}{2\gamma^{2}d^{2}}<c\left(\frac{h}{d}\right)^{4/3},\qquad c\simeq1.6\label{eq:Spin_reversal_condition_diss-less}
\end{equation}
at $h/d\ll1$ and $\alpha=0$.

\begin{figure}
\centering\includegraphics[width=8cm]{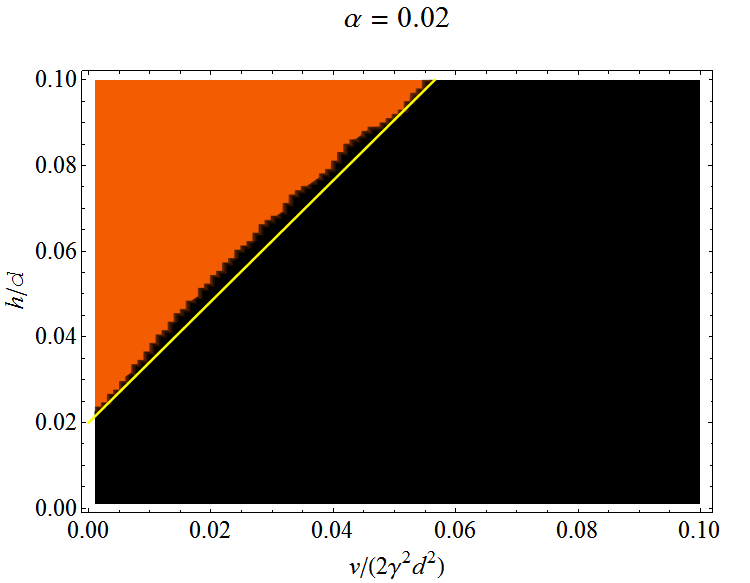}

\caption{Phase diagram of the magnetization reversal for $\alpha=0.02$. Yellow
line: Eq. (\ref{eq:slow-reversal-condition-d^2}).}

\label{Fig-PD-indep-linear-h_alpha=00003D00003D0.02_100x100} 
\end{figure}

Phase diagram of the magnetization reversal in the damped case $\alpha=0.02$
is shown in Fig. \ref{Fig-PD-indep-linear-h_alpha=00003D00003D0.02_100x100}.
It is similar to Fig. \ref{Fig-PD-indep-linear-h_alpha=00003D00003D0_100x100}
but there is a threshold for the magnetization reversal on $h$ and
the phase-boundary line goes linearly at small $v$. Computations
for different values of $\alpha$ suggest that the reversal requires
$h/d>\alpha$.

\begin{figure}
\centering\includegraphics[width=8cm]{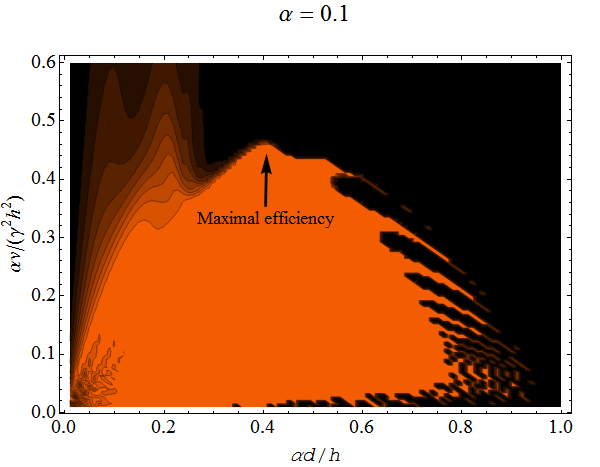}
\caption{Efficiency-type phase diagram of the magnetization reversal for $\alpha=0.1$.}

\label{Fig-PD-efficiency-alpha=00003D00003D0.1_100x100} 
\end{figure}

\begin{figure}
\centering\includegraphics[width=8cm]{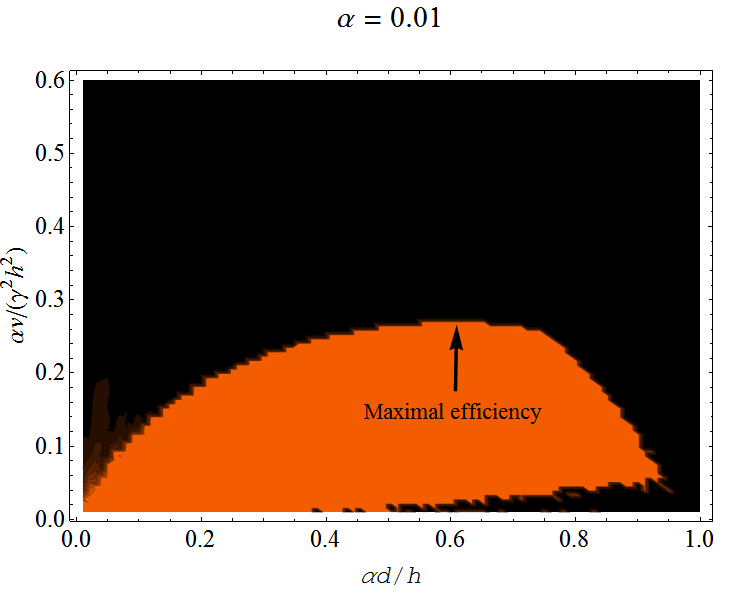}

\caption{Efficiency-type phase diagram of the magnetization reversal for $\alpha=0.01$.}

\label{Fig-PD-efficiency-alpha=00003D00003D0.01_100x100} 
\end{figure}

One can compute other types of phase diagrams for the magnetization
reversal that show a compact reversal region and the whole boundary
line. The most useful of these phase diagrams uses the parameters
$\left(\alpha d/h,\,\alpha v/(\gamma^{2}h^{2})\right)$. Indeed, the
area of the magnetization reversal is the compact region $0<\alpha d/h<1$
and $v/h^{2}$ is inversely proportional to the energy of the ac field
injected during the time of the reversal by the linear frequency sweep
\begin{equation}
t_{\mathrm{rev}}^{(\mathrm{linear})}=\frac{2\gamma d}{v}.\label{t-l}
\end{equation}
The maximum of $v/h^{2}$ corresponds to the minimal injected energy
and thus to the maximal efficiency of the reversal. Figs. \ref{Fig-PD-efficiency-alpha=00003D00003D0.1_100x100}
and \ref{Fig-PD-efficiency-alpha=00003D00003D0.01_100x100} show that
the maximal efficiency of the time-linear frequency sweep corresponds
to $\alpha d/h\approx0.5$. Also in these figures one can see that
there is no reversal if the sweep rate is too low, especially for
low ac fields on the right side.

\section{Analytical}

\label{sec:Analytical}

Analytical investigation of the magnetization reversal is more convenient
in spherical coordinates 
\begin{equation}
s_{z}=\cos\theta,\qquad s_{x}=\sin\theta\cos\varphi,\qquad s_{y}=\sin\theta\sin\varphi.
\end{equation}
After neglecting the ac field in the dissipation term, Eq. (\ref{eq:LLE-Rotating_frame})
becomes 
\begin{eqnarray}
\dot{\theta} & = & \gamma h\sin\varphi-\alpha\gamma d\sin2\theta\label{eq:theta(t)-Eq}\\
\dot{\varphi} & = & -2\gamma d\cos\theta-\omega\left(t\right)+\gamma h\cos\varphi\cot\theta.\label{eq:phi(t)-Eq}
\end{eqnarray}

\subsection{Linear frequency sweep}

\label{analytical-linear}

In the case of a linear frequency sweep, Eq. (\ref{eq:omega_Def}),
one can rewrite the equation of motion for the spin in terms of the
dimensionless time variable 
\begin{equation}
\tau=vt/(2\gamma d).\label{eq:tau-Def}
\end{equation}
The resulting equation of motion has the form 
\begin{eqnarray}
d\theta/d\tau & = & b\sin\varphi-\alpha a\sin2\theta\label{eq:theta-Eq}\\
d\varphi/d\tau & = & -2a\left(\cos\theta-\tau\right)+b\cos\varphi\cot\theta,\label{eq:phi-Eq}
\end{eqnarray}
where 
\begin{equation}
a\equiv\frac{2\gamma^{2}d^{2}}{v},\qquad b\equiv\frac{2\gamma^{2}dh}{v}\label{eq:a-b-Def}
\end{equation}
characterise the sweep rate. Another important parameter is 
\begin{equation}
A=\alpha d/h.\label{eq:A-Def}
\end{equation}

\begin{figure}
\centering\includegraphics[width=8cm]{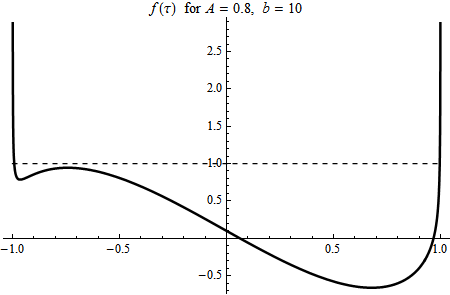}

\caption{$f(\tau)$ of Eq. (\ref{eq:Max-Eq}).}

\label{Fig-f(tau)} 
\end{figure}

Since $a$ is a large parameter, phase locking of the magnetic moment
to the ac field and thus efficient reversal requires $\cos\theta\cong\tau$
in the reversal region $|\tau|<1$. If $\cos\theta$ only slightly
deviates from this form, this causes strong oscillations of $\varphi$
and thus the breakdown of the phase locking. Setting $\cos\theta=\tau$,
from Eq. (\ref{eq:theta-Eq}) one obtains the phase-locking condition
for $\varphi$ in the form 
\begin{equation}
\sin\varphi=\frac{1}{b}\frac{d\theta}{d\tau}+A\sin2\theta.\label{eq:sin(phi)_via_theta}
\end{equation}
The term on the left of this formula is proportional to the torque
acting on the spin from the ac field. This torque has to ensure temporal
change of $\theta$ (i.e., reversal) and compensate for the dissipative
torque that is acting toward potential wells. One can see that damping
hampers climbing the barrier by the magnetic moment. The maximal damping
torque is realized at $\theta=3\pi/4$, where $\sin2\theta=-1$. Since
the reversal implies $d\theta/d\tau<0,$ it is clear that for $A>1$
the ac torque cannot overcome the damping torque. Thus, the necessary
condition for the magnetization reversal is 
\begin{equation}
A<1,\label{eq:h_gtr_alphad-condition}
\end{equation}
while the more restricting sufficient condition requires that the
right-hand side of Eq. (\ref{eq:sin(phi)_via_theta}) does not drop
below -1 for all $\tau$. The latter requires the frequency sweep
rate to be not too fast. Using $\cos\theta=\tau$, one can rewrite
this condition in the form 
\begin{equation}
\max f(\tau)<1,\qquad f(\tau)\equiv-2A\tau\sqrt{1-\tau^{2}}+\frac{1}{b\sqrt{1-\tau^{2}}}.\label{eq:Max-Eq}
\end{equation}
Because of the inertial term, $f(\tau)$ shown in Fig. \ref{Fig-f(tau)}
diverges at the borders of the reversal interval. This is, however,
an artefact of neglecting the rounding of the dependence $s_{z}(\tau)$
at $\tau=\pm1$ because of the finite value of $h$. When this effect
is taken into account, there are maxima around $\tau=\pm1$ instead
of divergences. Thus, spin reversal can break down either because
of the inertial effect near $\tau=-1$ or because of the effect of
dissipation near $\theta=3\pi/4$, i.e., $\tau=-1/\sqrt{2}$, depending
on which one occurs at a smaller sweep rate.

Let us first consider the dissipative breakdown of the magnetization
reversal for $A$ slightly below 1 that happens at a small sweep rate.
In this case $1/b\propto v$ is small and the second term in $f(\tau)$
in Eq. (\ref{eq:Max-Eq}) is a perturbation. Thus the value of this
term can be taken at the unperturbed dissipative maximum $\tau=-1/\sqrt{2}$.
Using $-2\tau\sqrt{1-\tau^{2}}=1$ and $\sqrt{1-\tau^{2}}=1/\sqrt{2}$,
one obtains the reversal condition 
\begin{equation}
A+\frac{\sqrt{2}}{b}<1.\label{eq:Slow-reversal-condition-A-eps}
\end{equation}
This can be rewritten in real units as 
\begin{equation}
\frac{v}{2\gamma^{2}d^{2}}<\frac{1}{\sqrt{2}}\left(\frac{h}{d}-\alpha\right)\label{eq:slow-reversal-condition-d^2}
\end{equation}
and it is in a reasonable agreement with the numerical results in
Fig. \ref{Fig-PD-indep-linear-h_alpha=00003D00003D0.02_100x100}.
Although this expression formally survives in the dissipationless
limit $\alpha\rightarrow0$, it becomes inapplicable in this limit.
Here the breakdown of spin reversal is due to the inertial effect.

To investigate the latter, one needs a more accurate approximation
for $f(\tau)$ in Eq. (\ref{eq:Max-Eq}) near $\tau=-1$ that tranforms
divergence into a maximum. This can be done by solving Eq. (\ref{eq:Ham_max_Eq})
although it is difficult to do it analytically in general. Instead,
since the maximum should be close to $\tau=-1$, we can solve this
equation exactly at $\tau=-1$, which is much easier. A perturbative
solution for $h/d\ll1$ yields 
\begin{equation}
d\dot{s}/d\tau\cong2/3,\qquad\sin\theta\cong\left(h/d\right)^{1/3}.
\end{equation}
and then one obtains 
\begin{equation}
-\frac{d\theta}{d\tau}=\frac{1}{\sin\theta}\frac{d\cos\theta}{d\tau}=\frac{2}{3}\left(\frac{d}{h}\right)^{1/3}.
\end{equation}
Replacing in Eq. (\ref{eq:Max-Eq}) $1/\sqrt{1-\tau^{2}}$ by this
result and using Eq. (\ref{eq:a-b-Def}), one obtains the dissipationless
reversal condition 
\begin{equation}
\frac{v}{2\gamma^{2}d^{2}}<\frac{3}{2}\left(\frac{h}{d}\right)^{4/3},\label{eq:Slow-sweep-cond-diss-less}
\end{equation}
in a reasonable agreement with the numerical result, Eq. (\ref{eq:Spin_reversal_condition_diss-less}).

The combined reversal condition obtained from Eqs. (\ref{eq:slow-reversal-condition-d^2})
and (\ref{eq:Spin_reversal_condition_diss-less}) is thus 
\begin{equation}
\frac{v}{2\gamma^{2}d^{2}}<\min\left\{ \frac{1}{\sqrt{2}}\left(\frac{h}{d}-\alpha\right),\frac{3}{2}\left(\frac{h}{d}\right)^{4/3}\right\} .\label{eq:slow-reversal-condition-d^2-1}
\end{equation}

Let us shortly discuss the stability of our spin-reversal solutions
that, in the laboratory frame, is the stability of phase locking between
the spin and the ac field at slow frequency sweep. Linearizing Eqs.
(\ref{eq:theta(t)-Eq}) and (\ref{eq:phi(t)-Eq}) around the static
solution ($\theta,\varphi)$ at a fixed time, one obtains the deviation
($\delta\theta,\delta\varphi)\propto e^{\lambda t}$. For the orientations
closer to the wells, $3\pi/4<\theta\leq\pi$ and $0\leq\theta<\pi/4$,
one has $\lambda<0$ and phase locking is stable. However, for the
orientations closer to the barrier, $\pi/4<\theta<3\pi/4$, one has
$\lambda>0$ and phase locking is unstable. Thus the barrier has to
be crossed fast enough during reversal before the instability develops.
Considering the process quasi-statically, one can write 
\begin{equation}
\left(\delta\theta,\delta\varphi\right)\sim\exp\left[\intop_{t_{0}}^{t}dt'\lambda(t')\right]
\end{equation}
and use the stability criterion $\intop_{t_{0}}^{0}dt\lambda(t)<1$,
where $t_{0}$ is the time of entering the instability region and
the top of the barrier is reached at $t=0.$ After some algebra one
arrives at the stability criterium 
\begin{equation}
\frac{\alpha}{3\sqrt{2}}<\frac{v}{2\gamma^{2}d^{2}}.
\end{equation}
A boundary of this kind is seen in Figs. \ref{Fig-PD-efficiency-alpha=00003D00003D0.1_100x100}
and \ref{Fig-PD-efficiency-alpha=00003D00003D0.01_100x100} close
to the bottom.

\subsection{Optimal frequency sweep}

\label{optimal}

\begin{figure}
\centering\includegraphics[width=8cm]{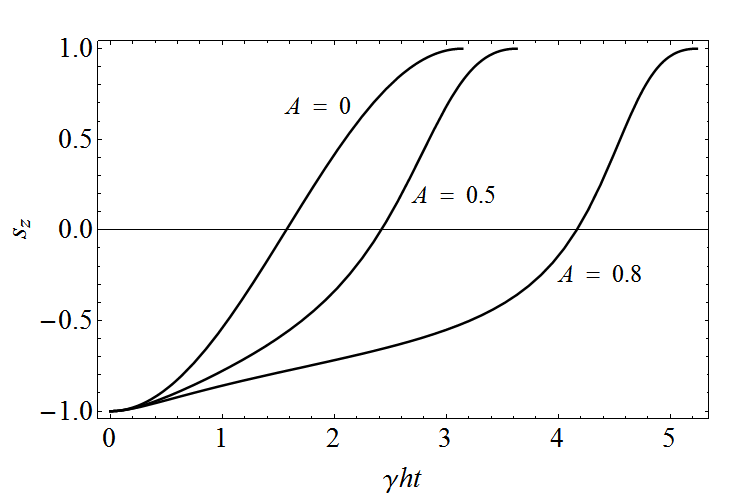} \caption{Optimal magnetization reversal for different values of $A=\alpha d/h$.}

\label{Fig-optimal_sweep} 
\end{figure}

The magnetization reversal can be optimized by applying a time-nonlinear
frequency sweep. Among all possible cases the so-called ``optimal
sweep'' stands out as a rotation of the magnetic moment in one plane
(in the rotating frame) with $\varphi=-\pi/2$, that is with the moment
being always perpendicular to the ac field. It is easy to see that
this provides the maximal torque on the magnetization during the reversal.
With

\begin{equation}
\omega\left(t\right)=-2\gamma d\cos\theta\label{eq:omega(t)-optimal_sweep}
\end{equation}
Eq. (\ref{eq:phi(t)-Eq}) self-consistently yields $\dot{\varphi}=0$.
Then Eq. (\ref{eq:theta(t)-Eq}) takes the form 
\begin{equation}
\dot{\theta}=-\gamma h-\alpha\gamma d\sin2\theta.
\end{equation}
Integrating this equation with the initial condition $\theta(0)=\pi$,
one obtains 
\begin{equation}
\tan\theta=\frac{-\sin\left(\tilde{t}\right)}{\cos\left(\tilde{t}-\arcsin A\right)},\label{eq:tan_theta-optimal}
\end{equation}
where $\tilde{t}\equiv\sqrt{1-A^{2}}\gamma ht$ and $A$ is defined
by Eq. (\ref{eq:A-Def}). After some algebra the expression for the
optimal sweep can be transformed to the most convenient form: 
\begin{equation}
s_{z}=\cos\theta=-\frac{\cos\left(\tilde{t}-\arcsin A\right)}{\sqrt{1-A\cos\left(2\tilde{t}+\arccos A\right)}}\label{eq:s_z(t)-optimal}
\end{equation}
illustrated in Fig. \ref{Fig-optimal_sweep}. Together with Eq. (\ref{eq:omega(t)-optimal_sweep})
it gives a non-linear time dependence of the frequency of the ac field
that provides the fastest reversal of the magnetic moment. This exact
result, that generalizes the result of Ref. \onlinecite{Rivkin-APL2006}
obtained in the absence of damping, must have important practical
applications. 

In the dissipationless case, $A\rightarrow0$, the optimal magnetization
reversal is described by a pure cosine function that is a Rabi precession
of the magnetic moment around the ac field. In the general case, the
reversal is mainly due to the cos term in the numerator, whereas the
denominator only affects the shape of the switching curve, making
it non-sinusoidal in the presence of damping.

Eq. (\ref{eq:s_z(t)-optimal}) and Fig. \ref{Fig-optimal_sweep} describe
the optimal reversal during the time 
\begin{equation}
t_{\mathrm{rev}}=\frac{\pi}{\sqrt{1-A^{2}}\gamma h},\qquad A\equiv\frac{\alpha d}{h}<1\label{eq:t_rev-optimal_sweep}
\end{equation}
It is instructive to compare this time with the time of the magnetization
reversal for the linear sweep, defined by Eq. (\ref{t-l}) (notice
that the total time of the process may be longer). In the dissipationless
case, the maximal sweep speed is given by Eq. (\ref{eq:Spin_reversal_condition_diss-less}).
For that speed one obtains the minimal reversal time 
\begin{equation}
t_{\mathrm{rev}}^{(\mathrm{linear})}=\frac{4}{3\gamma h}\left(\frac{d}{h}\right)^{1/3}
\end{equation}
that is longer than the time given by Eq. (\ref{eq:t_rev-optimal_sweep})
with $A=0$. In the dissipative case near $A=1$, the maximal sweep
rate for the linear sweep follows from Eq. (\ref{eq:slow-reversal-condition-d^2}).
This yields 
\begin{equation}
t_{\mathrm{rev}}^{(\mathrm{linear})}=\frac{2\sqrt{2}}{\gamma h}\frac{1}{1-A}.\label{eq:t_rev-linear_sweep}
\end{equation}
For $\left|1-A\right|\ll1$ this is also is much longer than the time
given by Eq. (\ref{eq:t_rev-optimal_sweep}). In the relevant region
$A\sim1$ Eqs. (\ref{eq:t_rev-optimal_sweep}) and (\ref{eq:t_rev-linear_sweep})
are comparable. However, one has to keep in mind that the linear frequency
sweep has to begin with a frequency beyond the resonance range, so
that the actual reversal time of the linear sweep is somewhat longer
than above.

The total energy input of the ac power needed for the reversal satisfies
\begin{equation}
E\propto h^{2}t_{\mathrm{rev}}.
\end{equation}
For the optimal sweep one has 
\begin{equation}
E\propto\frac{h}{\sqrt{1-A^{2}}}=\frac{h^{2}}{\sqrt{h^{2}-\left(\alpha d\right)^{2}}}.
\end{equation}
The minimum of this function, $2\alpha d$, is achieved at 
\begin{equation}
h=\sqrt{2}\alpha d
\end{equation}
(that is at $A=\alpha d/h=1/\sqrt{2}$).

For the linear sweep one has 
\begin{equation}
E^{(\mathrm{linear})}\propto h^{2}t_{\mathrm{rev}}^{(\mathrm{linear})}=\frac{h}{1-A}=\frac{h^{2}}{h-\alpha d}.
\end{equation}
The minimum of this function, $4\alpha d$, is achieved at 
\begin{equation}
h=2\alpha d
\end{equation}
(that is, at $A=\alpha d/h=0.5$).

We see that the magnetization reversal by the optimal sweep requires
both a smaller amplitude of the ac field and a smaller total energy
input, as compared to the linear sweep. In both cases the injected
energy at the maximal efficiency is proportional to $\alpha$ and
thus the efficiency itself is inversely proportional to $\alpha$.
(The latter was multiplied by $\alpha$ in Figs. \ref{Fig-PD-efficiency-alpha=00003D00003D0.1_100x100}
and \ref{Fig-PD-efficiency-alpha=00003D00003D0.01_100x100} to make
them approximately scale with $\alpha$.)

\section{ Reversal of the magnetization by Josephson currents}

\label{JJ}

\begin{figure}
\centering\includegraphics[width=65mm]{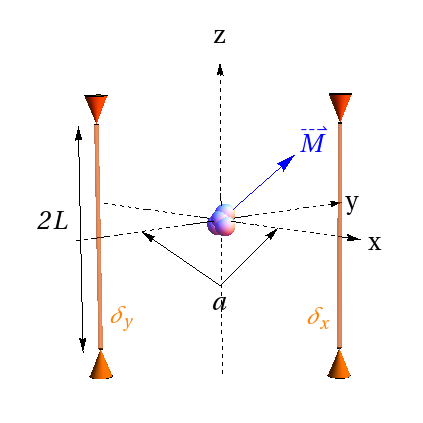}
\caption{Geometry used in the model: Nanomagnet makes the right angle with
two parallel superconducting weak links of length $2L$ at a distance
$a$ from the magnet.}

\label{two-JJ-one-nanomagnet-model} 
\end{figure}

Pointed switching of the magnetization of a nanomagnet by the ac field
of varying frequency may be achieved by coupling the magnet to a weak
superconducting link \cite{LC-PRB2010}. Advantage of this method
consists of the possibility to control the time dependence of the
frequency by the voltage across the link, $V(t)$. As has been discussed
in the Introduction the most effective switching occurs when the ac
field is circularly polarized. This requires two weak links shown
in Fig. \ref{two-JJ-one-nanomagnet-model}. In addition to the previous
problem one should now take into account the back effect of the magnet
on the weak links. As we shall see below, our results for the optimal
sweep permit generalization that provides exact time dependences of
voltages on the two links needed to obtain full reversal of the magnetization.

Each superconducting weak link interacting with the magnet contributes
the term 
\begin{equation}
{\cal E}_{J}=-E_{J}\cos\left[\delta-\frac{2\pi}{\Phi_{0}}\int_{1}^{2}d{\bf r}\cdot{\bf A}({\bf r},t)\right]\label{E-J}
\end{equation}
to the total energy. Here $E_{J}=\hbar I_{c}/(2e)$ is the Josephson
energy of the link, with $I_{c}$ being the critical current. The
argument of cosine is the gauge invariant phase that consists of two
contributions. The first contribution satisfying $\dot{\delta}=2eV(t)/\hbar$
comes from the voltage across the link, while the second contribution
is due to the vector potential of the magnet integrated between the
terminals of the link; $\Phi_{0}=2\pi\hbar c/(2e)$ being the flux
quantum. Following the footsteps of Ref.\ \onlinecite{LC-PRB2010},
it is easy to show that interaction of the magnet with two Josephson
junctions leads to a modified expression (\ref{eq:h_eff_Def}) for
the effective field, 
\begin{equation}
{\bf H}_{\mathrm{eff}}=2ds_{z}{\bf e}_{z}+h_{J}[\sin(\delta_{y}-ks_{x}){\bf e}_{x}+\sin(\delta_{x}-ks_{y}){\bf e}_{y}]\label{HeffJ}
\end{equation}
where $h_{J}=kE_{J}/\left(M_{s}V\right)$ is the amplitude of the
ac magnetic field created by the junction at the position of the nanomagnet,
\begin{equation}
\frac{d\delta_{x,y}}{dt}=\frac{2eV_{x,y}}{\hbar}\,,\label{dotPhixy}
\end{equation}
$V_{x,y}$ are the voltages across the junctions, and
\begin{equation}
k=\frac{4\pi M_{s}V}{a\Phi_{0}}\frac{L}{\sqrt{L^{2}+a^{2}}}
\end{equation}
 is a dimensionless spin-feedback coupling coefficient.

\begin{figure}
\centering\includegraphics[width=8cm]{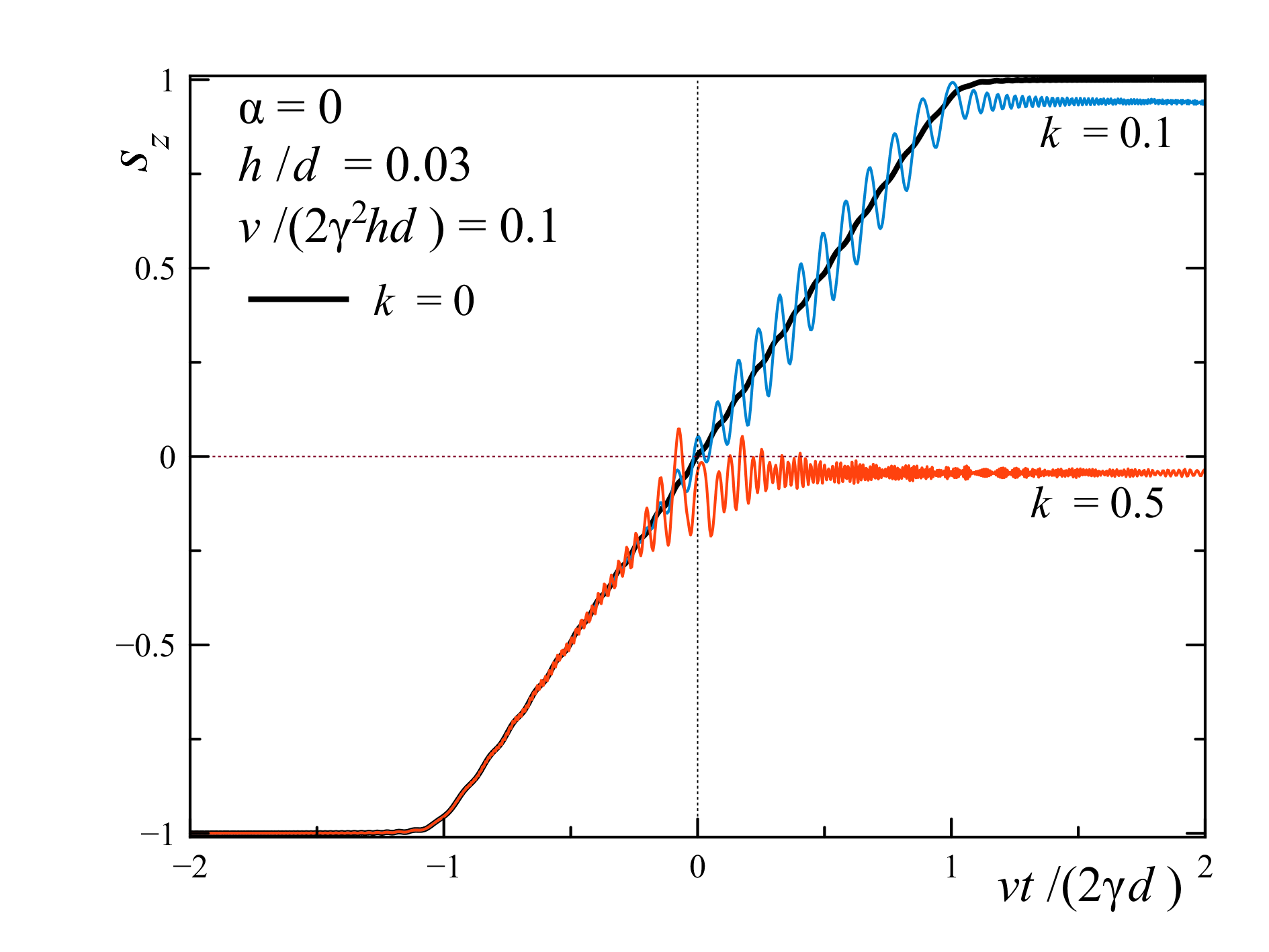}

\caption{Detrimental influence of the magnetization feedback on the Josephson
junction in the case of linear frequency sweep and zero damping.}

\label{Fig-spin_reversa-JJ-sz_alpha=00003D0}
\end{figure}

\begin{figure}
\centering\includegraphics[width=8cm]{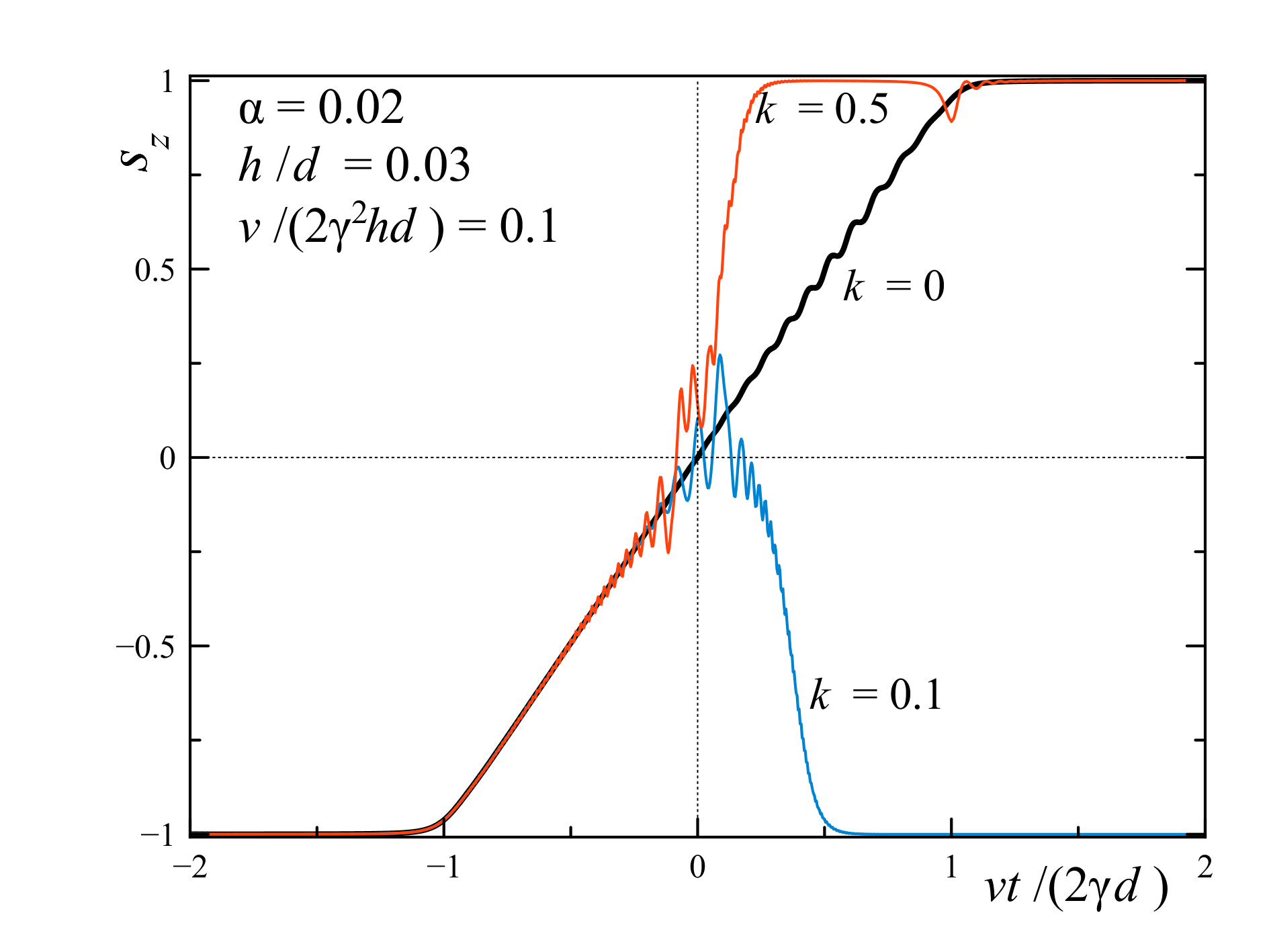}

\caption{Influence of the magnetization feedback in the damped case with linear
sweep.}

\label{Fig-spin_reversa-JJ-sz_alpha=00003D0.02}
\end{figure}

For the time-linear voltage, dependence of the effective field in
Eq. (\ref{HeffJ}) on oscillating transverse spin components is detrimental
for reversal. This happens because the oscillating additions to the
otherwise time-smooth phases disturb phase locking between the magnetization
and the ac field and cause non-adiabaticity. Fig. \ref{Fig-spin_reversa-JJ-sz_alpha=00003D0}
shows that non-adiabaticity in the zero-damping case becomes pronounced
already for small values of the feedback coefficient $k$. In the
realistic damped case the negative influence of finite $k$ is even
stronger. The instability of the phase locking for $\pi/4<\theta<3\pi/4$
discussed at the end of Sec. \ref{analytical-linear} exponentially
increases the mismatch between the directions of the ac field and
the magnetization arizing because of the feedback. As a result, the
magnetization randomly lands in one of the two wells, as shown in
Fig. \ref{Fig-spin_reversa-JJ-sz_alpha=00003D0.02}.

\begin{figure}
\centering\includegraphics[width=85mm]{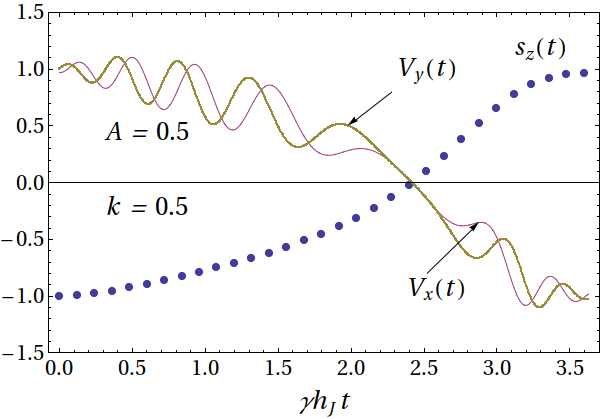} \caption{The optimal choice of $V_{x}$ (purple) and $V_{y}$(gold) across
the weak links at $A={k}=0.5$. Time dependence of $s_{z}$ is shown
by blue dots.}

\label{Mz-J} 
\end{figure}

In order to reduce the effect of the Josephson junctions on the magnet
to the effect of a circularly polarized field, one can require that
\begin{equation}
\delta_{x}-ks_{y}=\delta(t)\,,\qquad\delta_{y}-ks_{x}=\delta(t)+\frac{\pi}{2},\label{Phixy}
\end{equation}
where $\delta(t)$ is a phase. This allows one to use the results
of the previous section for the optimal sweep requiring $\dot{\delta}(t)=\omega\left(t\right)=-2\gamma d\cos\theta$.
For such a sweep, $s_{x,y}(t)$ in the laboratory frame are precessing
diring reversal as follows 
\begin{align}
s_{x}=\sin\theta(\tilde{t})\cos\delta(\tilde{t})=\frac{\sin(\tilde{t})\cos\delta(\tilde{t})}{\sqrt{1-A\cos(2\tilde{t}+\arccos A)}}\nonumber \\
s_{y}=\sin\theta(\tilde{t})\sin\delta(\tilde{t})=\frac{\sin(\tilde{t})\sin\delta(\tilde{t})}{\sqrt{1-A\cos(2\tilde{t}+\arccos A)}},\label{Mxy}
\end{align}
where $A=\alpha d/h_{J}$, $\tilde{t}=\sqrt{1-A^{2}}\gamma h_{J}t$,
and $\sin\theta(\tilde{t})$ was obtained by combining Eqs. (\ref{eq:tan_theta-optimal})
and (\ref{eq:s_z(t)-optimal}). The time-dependent phase is given
by 
\begin{equation}
\delta(t)=\int_{0}^{t}\omega(t')dt'=-\frac{2d/h_{J}}{\sqrt{1-A^{2}}}\int_{0}^{\tilde{t}}\cos\theta(\tilde{t}')d\tilde{t}'\,.
\end{equation}
with $\omega(t)$ defined by Eqs.\
(\ref{eq:omega(t)-optimal_sweep}) and $\theta(t)$ given by (\ref{eq:s_z(t)-optimal}).
In accordance with Eq.\ (\ref{Phixy}), the optimal time dependence
of the voltages across the junctions become 
\begin{equation}
V_{x}(t)=\frac{\hbar}{2e}[\omega(t)+k\dot{s}_{y}]\,,\qquad V_{y}(t)=\frac{\hbar}{2e}[\omega(t)+k\dot{s}_{x}].\label{Vxy}
\end{equation}
This dependence is shown in Fig.\
\ref{Mz-J}. The time dependence of $s_{z}$ in the figure is the
same as for the optimal freqency sweep in Fig.\
\ref{Fig-optimal_sweep}. Oscillations of the voltages are due to
the terms in Eq.\ (\ref{Vxy}) that depend on $s_{x,y}$. They are
weak as long as $k$ is small. Oscillations disappear in the limit
of $k\rightarrow0$, making $V_{x,y}$ in that limit to follow the
smooth time dependence of the optimal frequency sweep obtained in
the previous section. 

\begin{figure}
\centering\includegraphics[width=85mm]{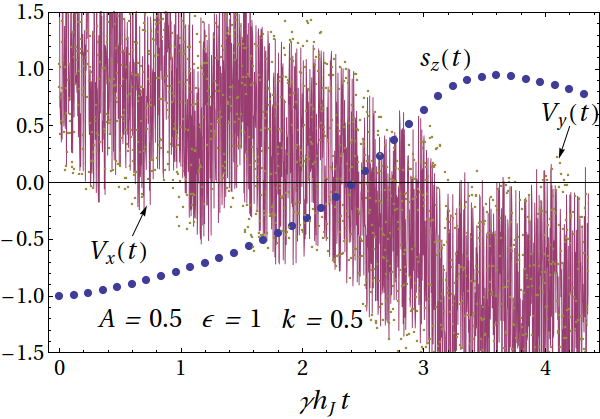} \caption{Optimal magnetization reversal in the presence of voltage noise in
the weak links.}

\label{Mz-J-disturb} 
\end{figure}

Having practical applications in mind, it is interesting to test the
stability of the reversal described by the above equations against
high-frequency voltage noise. This can be done by writing 
\begin{equation}
V'_{x,y}(t)=V_{x,y}(t)+\epsilon\frac{\hbar\gamma d}{e}F_{x,y}(t)
\end{equation}
with $F_{x,y}$ being uniform random functions of time between $-1$
and $+1$ and $\epsilon$ representing the relative strength of the
noise. Quite remarkably, as is illustrated in Fig.\
\ref{Mz-J-disturb}, the full magnetization reversal may occur even
in the presence of a strong noise. At $\epsilon=1$ this happens with
more than 0.99 probability. With less than 0.01 probability the magnetic
moment bounces back to $s_{z}=-1$ before it reaches $s_{z}=1$. This
can be traced to the fact that the high-frequency noise in most cases
averages out in the phase $\delta$ because the latter is proportional
to the time integral of the voltage. 

\section{Discussion and Conclusions}

\label{conclusion}

We have studied numerically and analytically a microwave-assisted
reversal of the magnetic moment of a single-domain magnetic particle.
We conclude from our studies that a circularly polarized ac field
that has specific time dependence of the frequency can be an effective
tool for switching the magnetization. The corresponding physical mechanism
consists of the resonant absorbtion of photons of the spin projection
that ensures the consistent change in the projection of the magnetic
moment. Emission of excitations by the magnetic moment inhibits this
process. In the micromagnetic theory it is described by the phenomenological
dimensionless small parameter $\alpha$ that can be independently
measured. In single-domain particles this parameter is usually greater
than in bulk materials and is typically of order $0.01-0.1$. \cite{Coffey}
The condition on the power of the ac field needed to overcome damping
and reverse the magnetization is $h>\alpha d$. For, e,g., the anisotropy
field $d$ of order $0.01$T and $\alpha$ of order $0.01$, one obtains
$h$ of order $0.0001$T for the amplitude of the ac field, which
is a reasonable value from the practical point of view.

We have studied linear and nonlinear time dependence of the frequency
of the ac field. It has been demonstrated that the linear case, $\omega=-vt$,
resembles the Landau-Zener problem. Magnetization reversal has been
demonstrated numerically and the phase diagrams have been obtained
that show the range of $v$, $h$, $d$, and $\alpha$ that provide
the reversal. They show that for the reversal to occur, the frequency
sweep must be sufficiently slow, but not too slow when the damping
is finite. The linear case has also been studied analytically. Condition
(\ref{eq:slow-reversal-condition-d^2-1}) has been obtained for the
upper bound on the frequency sweep rate. For the values of the parameters
used above, that upper bound is in the ballpark of $10^{7}$GHz/s.
The minimal reversal time for the time-linear sweep is of order $(\gamma h)^{-1}$.

We have also studied a time-nonlinear frequency sweep. Exact analytical
solution for $\omega(t)$ that provides the fastest reversal has been
obtained with account of damping. It is given by equations (\ref{eq:omega(t)-optimal_sweep})
and (\ref{eq:s_z(t)-optimal}). This finding may have important practical
application. We call this sweep the optimal sweep. It has been demonstrated
that, besides ensuring the fasted magnetization switch, it also pumps
less energy into the system as compared to the linear sweep. In both
cases the injected energy is proportional to $\alpha$.

Circularly polarized ac field can be generated by coupling a single-domain
particle electromagnetically to two weak superconducting links whose
phases are displaced by $\pi/2$ with respect to each other. One advantage
of such a system is that the time dependence of the frequency of the
ac field generated by the links can be controlled by voltage. This
problem has been studied by us with account of the back effect of
the magnetic moment on the links. Magnetization reversal has been
demonstrated numerically and analytical expressions have been derived
for the time dependence of the voltages across the links that provide
the fasted magnetization reversal. One remarkable property of this
system is weak dependence of the reversal dynamics on the voltage
noise.

\section{Acknowledgements}

D.G. is thankful to H. Kachkachi, N. Barros, and C. Thirion for useful
discussions of microwave-induced spin reversal.

This work has been supported by the U.S. Department of Energy through
Grant No. DE-FG02-93ER45487.



\end{document}